\documentclass[12pt,preprint]{aastex}

\newcommand{\myemail}{ajaunsen\@astro.uio.no}

\slugcomment{Submitted for publication in ApJ}

\shorttitle{GRB~070306: a highly extinguished afterglow}
\shortauthors{Jaunsen et al.}

\begin{document}


\title{GRB~070306: a highly extinguished afterglow\footnote{Based on observations made at the European Southern Observatory, Paranal, Chile under program 078.D-0416, 177.A-0591; with the Nordic Optical Telescope, operated on the island of La Palma jointly by Denmark, Finland, Iceland, Norway and Sweden in the Spanish Observatorio del Roque de los Muchachos of the Instituto de Astrof{\'i}sica de Canarias; with the William Hershel Telescope operated on the island of La Palma by the Isaac Newton Group in the Spanish Observatorio del Roque de los Muchachos of the Instituto de Astrof{\'i}sica de Canarias}}


\author{A.~O.~Jaunsen\altaffilmark{1}}
\affil{Institute of Theoretical Astrophysics, University of Oslo,
 PO Box 1029 Blindern, N-0315 Oslo, Norway}

\author{E.~Rol\altaffilmark{2}}
\affil{Department of Physics and Astronomy, University of Leicester, Leicester, LE1\, 7RH, U.K.}

\author{D.~J.~Watson, D.~Malesani, J.~P.~U.~Fynbo, B.~Milvang-Jensen, J.~Hjorth, P.~M.~Vreeswijk\altaffilmark{3}}
\affil{Dark Cosmology Centre, Niels Bohr Institute, University of Copenhagen, Juliane Maries Vej 30, DK-2100 Copenhagen ¯, Denmark}

\author{J.-E.~Ovaldsen\altaffilmark{1}}

\author{K.~Wiersema, N.~R.~Tanvir\altaffilmark{2}}

\author{J.~Gorosabel\altaffilmark{4}}
\affil{Instituto de Astrof{\'i}sica de Andaluc{\'i}a (IAA-CSIC), Apartado de Correos 3004, 18080 Granada, Spain}

\author{A.~J.~Levan\altaffilmark{5}}
\affil{Department of Physics, University of Warwick, Coventry, CV4\, 7AL, U.K.}

\author{M. Schirmer\altaffilmark{5}}
\affil{Isaac Newton Group of Telescopes (INT), Apartado de Correos 321, 38700 Santa Cruz de La Palma, Tenerife, Spain}

\and
\author{A.J.~Castro-Tirado \altaffilmark{4}}

%\and
%
%\author{R. J. Hanisch\altaffilmark{5}}
%\affil{Space Telescope Science Institute, Baltimore, MD 21218}


%\altaffiltext{1}{Visiting Astronomer, Cerro Tololo Inter-American Observatory.
%CTIO is operated by AURA, Inc.\ under contract to the National Science
%\altaffiltext{2}{Society of Fellows, Harvard University.}
%\altaffiltext{3}{present address: Center for Astrophysics,
%    60 Garden Street, Cambridge, MA 02138}


\begin{abstract}
We report on the highly extinguished afterglow of GRB~070306 and the properties of the host galaxy. An optical afterglow was not detected at the location of the burst, but in near-infrared a doubling in brightness during the first night and later power-law decay in the $K$ band provided a clear detection of the afterglow.
The host galaxy is relatively bright, $R\sim22.8$. An optical low resolution spectrum revealed a largely featureless host galaxy continuum with a single emission line. Higher resolution follow-up spectroscopy shows this emission to be resolved and consisting of two peaks separated by 7 \AA, suggesting it to be [O\,{\scriptsize II}] at a redshift of $z=1.49594 \pm 0.00006$. The infrared color $H-K = 2$ directly reveals significant reddening. By modeling the optical/X-ray spectral energy distribution at $t = 1.38$~days with an extinguished synchrotron spectrum, we derive $A_V = 5.5 \pm 0.6$ mag. This is among the largest values ever measured for a GRB afterglow and visual extinctions exceeding unity are rare. The importance of early NIR observations is obvious and may soon provide a clearer view into the once elusive 'dark bursts'.
\end{abstract}


\keywords{dust, extinction --- cosmology: observations --- galaxies: redshift and starburst --- gamma rays: bursts}



\section{Introduction}

Long gamma-ray bursts (GRBs) are believed to be indicators of star formation due to their association with short lived massive stars \citep{Galama1998,Hjorth2003,Stanek2003}. These events should therefore offer great prospects for characterizing the star formation history of the Universe \citep{Wijers1998}, as GRBs can be detected (in gamma and X-rays) at all observable redshifts and through large columns of dust and gas. One potential objection to this is if GRBs are biased towards a special subset of massive stars, i.e.\ those of low metallicity or in peculiar circumstellar environments. On the other hand, if there is no dependency on metallicity then the nature of the host galaxies leads to the surprising conclusion that most stars are formed in dwarf galaxies \citep{Fynbo2007}.

In the pre-Swift era the detection of optical afterglows (OAs) was a requirement for sub-arcsec localizations. This was obtained for about 30\% of the GRBs due to effects such as observability, weather conditions etc. However, in a few cases stringent upper limits on the early (optical) afterglow could be obtained \citep[see e.g.\ ][]{Groot1998}, providing evidence for a sub-class of OA-less bursts dubbed "dark burst" \citep[see e.g.\ ][]{Fruchter1999,Lamb2000,Fynbo2001,Berger2002,Lazzati2002}. Today the situation is quite different as the majority of bursts ($>$90\%) are localized with the Swift XRT instrument \citep[][]{Gehrels2004,Burrows2005}. \citet{Jakobsson2004} proposed an unbiased definition of a dark burst by suggesting that the spectral slope between the optical and X-ray is $\beta_{OX} < 0.5$. As many as 25\% of the Swift bursts fulfill this criterion \citep{Fynbo2007}. Throughout the paper we adopt this definition of a dark burst. There are several effects that may cause an afterglow to be classified as dark, but among the most obvious and possibly dominant effects are dust obscuration and high redshift \citep[see also][]{Roming2006}.

Although the afterglows of dark bursts do not all go by undetected, there is a sizeable sub-class of bursts for which we do not have much information and these may consequently affect the luminosity function and statistics of GRBs and their host galaxies significantly \citep[see e.g.\ ][who reported on a dark burst with significant extinction]{Rol2007}. There are also a few systems which do not resemble the general population of GRBs \citep{Perley2008}.

In this paper we present the detection of the NIR afterglow of GRB~070306 which is characterized as a dark burst. The most likely interpretation is that the afterglow is heavily dust obscured and the level of extinction provides a unique opportunity to study the local dust properties of the progenitor.
The outline of this paper is as follows: in Section~\ref{sec:obs} we describe the Swift X-ray data, and our optical and near-infrared observations of GRB~070306. The host and the spectral energy distribution (SED) template fitting is described in Section~\ref{sec:host}, the afterglow and the extinction curve fits are discussed in Section~\ref{sec:afterglow}, while the results are summarized and discussed in Section~\ref{sec:summary}.

Throughout the paper we assume a cosmology with $H_0 = 70 \ {\rm km}\ {\rm s}^{-1} {\rm Mpc}^{-1}$, $\Omega_M = 0.3$ and $\Omega_\Lambda = 0.7$. The afterglow is characterized by a standard temporal and spectral power-law on the form, $F \propto t^{-\alpha} \nu^{-\beta}$. All quoted errors are 1-sigma (68\%), unless noted differently.

\section{Observations and reductions}
\label{sec:obs}

GRB~070306 was discovered by \textit{Swift} at ($t_0$) 16:44:28.0 UT on 2007 March 06 \citep{gcnr38.2}. The burst had a $T_{90} = 210 \pm 10$\,s \citep{gcnc6173}. Swift slewed immediately and XRT observations began at $t_0 +  153$\,s and with UVOT at $t_0 + 162$\,s. An X-ray afterglow was identified and reported by \citet{gcnc6172}.

\subsection{X-ray}
\label{sec:xrt}

The Swift-XRT lightcurve shows a typical plateau phase, ending at approximately 35\,ks after the BAT trigger. To produce the X-ray spectrum at the time of the optical observations, the PC events after the plateau phase (35--266\,ks) were reduced in a standard way using the most recent calibration data and normalised to the count rate at the time of the optical observations ($t_0 + 1.38$\,days). A power law with fixed Galactic and variable absorption at the redshift of the host ($z=1.4959$, see sections \ref{sec:host} and \ref{sec:afterglow}) fits the data well ($\chi^2/{\rm dof} = 27.1/31$). The best-fit photon index $\Gamma=2.2\pm0.1$ ($\beta_X=1.2\pm0.1$); the excess absorption $N_{\rm H} = (2.6\pm0.4)\times10^{22}$\,cm$^{-2}$ for the equivalent hydrogen column density (assuming solar abundances), in excess of the Galactic value \citep[$3 \times 10^{20}$\, cm$^{-2}$, ][]{Dickey1990} corresponding to $(0.4\pm0.04)\times10^{22}$\, cm$^{-2}$ at $z=0$. We note that the fit with excess absorption at $z=0$ is poor. For sub-solar ($Z=0.1 Z_\odot$) metallicity we find $N_{\rm H} = (20\pm3)\times10^{22}$\,cm$^{-2}$. 
There is no evidence for a change in the spectral parameters between the plateau and post-plateau phases: the best fit absorption and spectral slope parameters were consistent to within one sigma.

The optical to X-ray spectral slope can be estimated at Mar 07.00 UT by using the specific 3\,keV flux at this time ($0.79\,\mu$Jy) and the estimated upper limit on the afterglow from the $I$ band NOT image at Mar 06.95 UT ($I>22$ mag(Vega), or $F_\nu < 4.1\ \mu$Jy). This gives an upper limit of $\beta_{OX} < 0.23$. This is, hence, a dark burst according to the definition of \citet{Jakobsson2004}, which requires that $\beta_{\rm OX} < 0.5$. 

\subsection{Optical and near-infrared}
\label{sec:optnirobs}

We carried out early follow up to localize the afterglow in the optical using the 2.5\,m Nordic Optical Telescope (NOT) and ALFOSC instrument and in the near-infrared using the 4m William Herschel Telescope (WHT) and LIRIS instrument on Observatorio del Roque de los Muchachos. Observing conditions were good, but the Moon was close to being full. Imaging and photometry data from the SDSS \citep{gcnc6170} revealed an object close to the detection limit within the error circle. 
\citet{gcnc6174} reported the detection of a fairly bright near-infrared (NIR) afterglow detected 3.3\,hr after the burst which had brightened by nearly a factor two 4.7\,hr later \citep{gcnc6176}. The NIR afterglow was consistent with the location of the optical SDSS detection. Optical follow up detected an object consistent in location and brightness with the object seen in the SDSS data \citep{gcnc6178}. The NIR follow up was continued the following night when we triggered Target-of-Opportunity (ToO) observations with VLT/ISAAC under program 078.D-0416(E) (PI: Vreeswijk). A FORS 300V optical spectrum was obtained (program ID 078.D-0416(B)) at the position of the NIR afterglow and optical SDSS detection in order to secure a redshift measurement. A second ISAAC epoch was obtained on the following night to further characterize the NIR afterglow. Unfortunately, continued follow up was not possible as all available time for this run had been exhausted. Finally, late time NIR observations to secure estimates of the host were obtained with the WHT/LIRIS in the end of April and a $K$ band point in January 2008 as part of a GRB host program at VLT/ISAAC (program ID 177.A-0591(Q), PI: J. Hjorth).

Using our best optical $R$ band image and 30 objects from the USNO B1.0 catalog we determine the location of the host galaxy to be ($\alpha$, $\delta$) = (09$^h$52$^m$23.30$^s \pm 0\farcs20$, $+10^{\circ}28^\prime55\farcs5 \pm 0\farcs12$). Images of the GRB afterglow and host in $R$ and $K$ band are shown in Figs.~\ref{fig:grbRimage} \& \ref{fig:grbimage}.
The afterglow in the $K$ band has the same location within the errors, although a small offset of not more than a pixel ($0\farcs15$) can be seen in Fig.~\ref{fig:grbimage}.  The host galaxy appears point-like in most of the filters, with the exception of the $K$ band, where the deeper image reveals some faint extended emission.

\subsubsection{Imaging}
\label{sec:phot}

The optical data were processed in the standard way using IRAF\footnote{Image Reduction and Analysis Facility (IRAF), a software system distributed by the National Optical Astronomy Observatories (NOAO).} and packages therein. The near-infrared data were processed using a combination of observatory pipelines (Eclipse, THELI) and IDL. Absolute photometric calibration was performed on the best available imaging data ($R$ on May 7, $I$ on Mar 12 and NIR on Mar 9) using SExtractor \citep{sextractor} and a fixed aperture of 3\farcs0 radius. The zero point was established from standard star observations for the NIR (ISAAC) data, while for the FORS and NOT (optical) data we used six field stars from the SDSS data to photometrically calibrate our $R$ and $I$ bands using the color transformations of \citet{Jester2005}. The remaining observations were calibrated by offsetting the photometry relative to these reference images.
The results are given in Table~\ref{tab:obslog}, and the reference images are denoted by daggers. The foreground reddening is estimated to $E(B-V) = 0.0275$ using the maps of \citet{Schlegel1998}, but the column containing afterglow (Vega) magnitudes given in Table~\ref{tab:obslog} have not been corrected for the Galactic extinction.

Optimal sensitivity to variability is achieved by degrading each frame to the worst PSF quality in a series  of images (for a given filter). The difference images are then inspected to identify residual emission at the afterglow position. We performed image subtraction using the ISIS package \citep{Alard1998} on all relevant imaging data. The reference frame for each filter (for registration and flux variability) was generally chosen as the latest epoch, but for $J$ and $H$ band these frames had a very low signal-to-noise and produced very poor image subtracted results. For these two series an earlier epoch had to be selected as a reference, although this was not ideal due to a possible afterglow contribution. The actual reference frames are indicated in Table~\ref{tab:obslog}. Prior to running ISIS, the images were registered and transformed using the IRAF \texttt{immatch} package in order to ensure a common coordinate system and smooth processing by ISIS. We also copied an isolated star (a pseudo-PSF) of similar flux as that of the afterglow to an empty region of sky near the location of the GRB so that it would remain in the field after the image subtraction (for comparison purposes). In those frames for which we detected a residual ($2\sigma$ level) at the location of the GRB we employed fixed aperture photometry (using SExtractor) and an aperture radius of $1\farcs5$ to measure the relative flux between the residual and the reference star. In the opposite case, in which no residual (at the $2\sigma$ level) could be detected, we estimated the detection limit by scaling our chosen PSF star by a decreasing amount and placing it at the location of the GRB afterglow/host. Image subtraction was then performed as described earlier and the magnitude difference was computed relative to the reference star (pseudo-PSF). This process was repeated until it could no longer be detected, determining the afterglow detection limit (or upper limit).

\begin{figure*}
  \centering
  \epsscale{1.0}
  \plotone{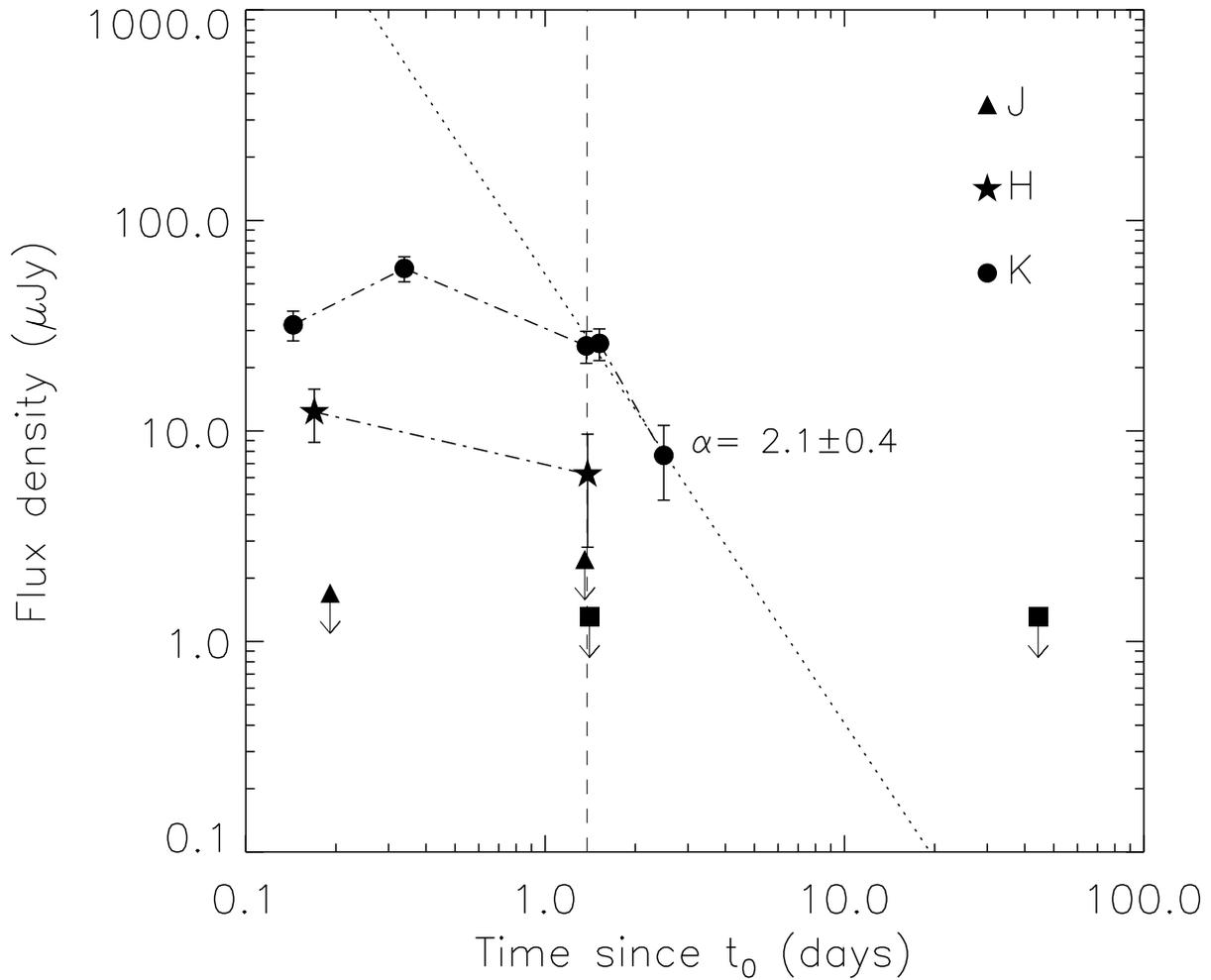}
  \figcaption[f1]{The multi-color lightcurve ($J$, $H$, $K$) of GRB~070306 and estimated late-time decay index, $\alpha = 2.1 \pm 0.4$. The vertical dashed line marks the simultaneity time, $t_0 + 1.38$\,days. Upper limits are denoted by an arrow. \label{fig:lightcurve}}
\end{figure*}

\begin{figure*}
  \centering
  \epsscale{1.0}
  \plotone{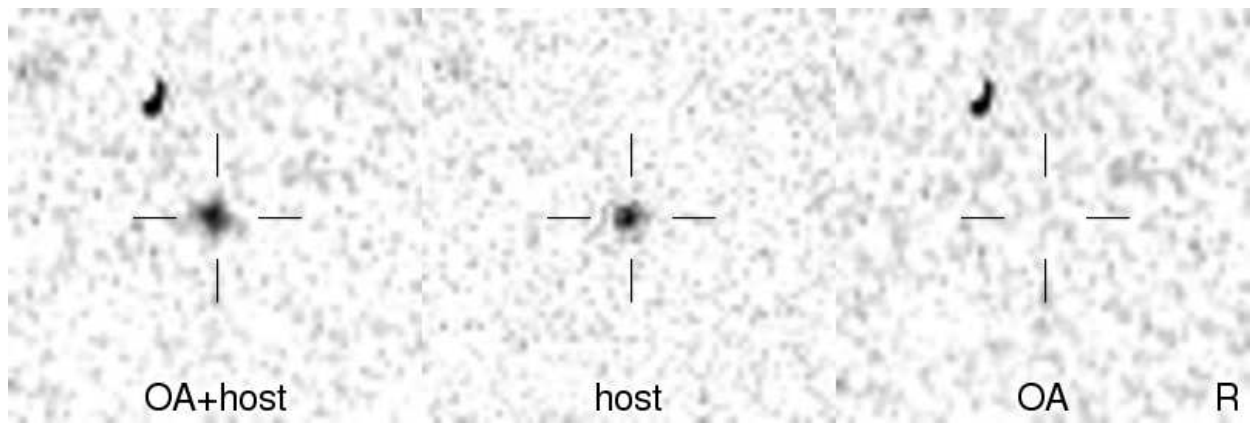}
  \figcaption[f2]{The $R$ band image at Mar 08.10 (left), combined Apr 20 and May 07 (middle) and the difference image (right) obtained using ISIS. No residual emission is detected and the estimated upper limit of the afterglow is $R > 23.5$ (estimated as described in Sec.~\ref{sec:phot}). North is up, and East is left. \label{fig:grbRimage}}
\end{figure*}

\begin{figure*}
  \centering
  \epsscale{1.0}
  \plotone{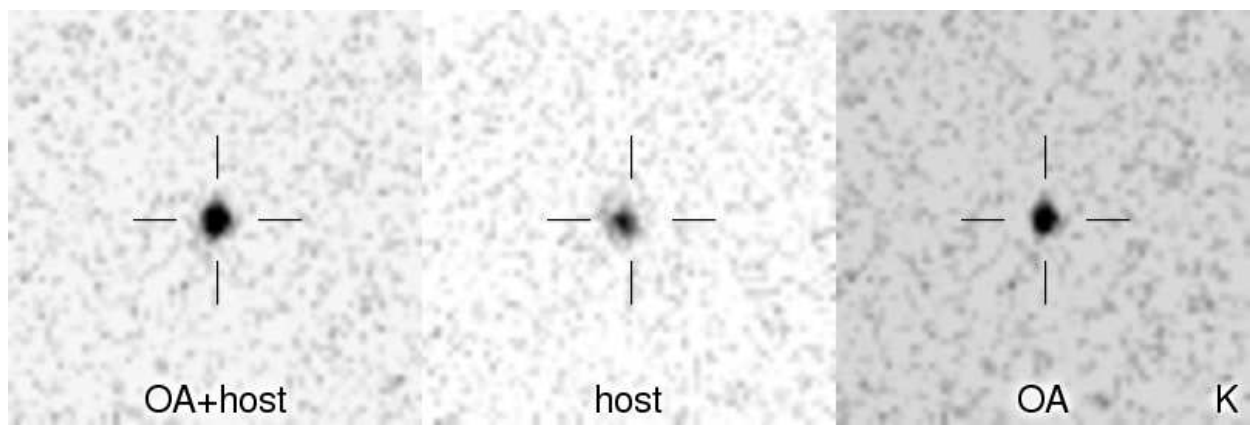}
  \figcaption[f3]{Image subtraction (ISIS) is used to detect variability and in this montage we show the $K$ band image of Mar 08.21 with contribution from the afterglow and host (left). In the middle frame the late-time (host) image from 2008 Jan 30.22 is shown, clearly revealing its extended nature. To the right the difference image reveals the significant contribution of the afterglow of GRB~070306 in the $K$ band. The images, which have been registered using field stars, show that the afterglow is centered on the brightest region in the host galaxy (although a small offset towards the north-west may be present on the order of a pixel, $0\farcs15$). North is up, and East is left. \label{fig:grbimage}}
\end{figure*}

The multi-color light curve ($J,H,K$ bands) is shown in Fig.~\ref{fig:lightcurve} and is characterized by an initial rise to peak during the first day in the $K$ band and possibly also in the $H$ band. The subsequent decline can be approximated by a power law, $F \propto t^{-\alpha}$. During the second and third days we fit the power law decline in the $K$ band with $\alpha = 2.1 \pm 0.4$ typical for post-break lightcurves \citep[e.g.\,][their Fig.~4]{Andersen2000}. No slope could be measured for the $J$ and $H$ bands. In the $H$ band the afterglow is clearly detected in the first two epochs, but the poorer image quality and lower signal-to-noise of the last epoch prevented us from setting meaningful limits on the earlier epochs and we were therefore forced to use the third epoch as a reference image. In addition, the images originate from two different instruments, which clearly does not help. Consequently, we are unable to detect an afterglow component at this epoch and may possibly overestimate the flux of the host (if there is a detectable contribution from the afterglow in this third epoch). We had similar problems with the $R$, $I$ and $J$ series, although the afterglow is not detected in these bands. In the $J$ band there may be a marginal signal in the first epoch (at $t_0 + 0.19$\,days), but this is not significant enough to be measured.

\placefigure{fig:lightcurve}

The $K$ band light curve displays a prominent brightening during the first day and although this has been observed in a few cases, this is a rare event \citep{Pedersen1998, Stanek2007, Guidorzi2007}. There have been a number of explanations for such phenomena ranging from some form of energy injection, dust echos, micro-lensing events to density discontinuities \citep{Tam2005,Lazzati2002.2}.

\subsubsection{Spectroscopy}
\label{sec:spec}

On 2007 March  8 (starting at 02:38 UTC) using the VLT UT1/FORS2 instrument with the 300V grism we obtained three spectra with a total exposure time of 5400\,s. The spectrum displayed a featureless continuum from 3700 to 9500 \AA, apart from a single emission line at $\sim$9306 {\AA}. This was attributed to [O\,{\scriptsize II}] $\lambda3727$ at a redshift of 1.5 on account that no other emission or absorption lines were detected \citep{gcnc6202}. The line significance is relatively high, but it is blended with the telluric skylines at 9306 {\AA} and onwards and with a resolution of about 12 {\AA} this prevented the line from being unambiguously identified. We have considered other potential line identifications such as H$\beta$, [O\,{\scriptsize III}] $\lambda5007$ or H$\alpha$, all of which are rejected on account that we expect to see additional and roughly similar bright lines in the wavelength range 6900-7140 \AA.

%
% 300V fluxave: flux=4.5e-17 erg/cm2/s/A  err: 1.1e-17
%

\placefigure{fig:300V_spec}

In an attempt to resolve the [O\,{\scriptsize II}] doublet (expected line separation is 7 {\AA}), we observed the host galaxy with the 1028z grism as part of our large program 177.A-0491(J) (PI: Hjorth). The observations were carried out in service mode on 2007 May 7 using VLT/FORS2 with a total exposure time of 2600\,s. The seeing was $0\farcs7$ and sky transparency was clear. The spectra were processed with special emphasis on obtaining a good sky subtraction. 
The issue is the fact that the skylines in these spectra are somewhat tilted, i.e.\ not perfectly aligned with the $y$--axis. In a traditional reduction the frames are first wavelength calibrated, which involves an interpolation of the original data values, and then sky subtracted. Since the steep flanks of the skylines are not sufficiently well sampled, the detailed shapes of the skylines in the interpolated frames vary from row to row, leading to systematic residuals in the sky subtracted frames. We dealt with this problem by using the method of \citet{Kelson2003}, as implemented in \citet{Milvang-Jensen2008}. 
%In this method, the sky is fitted and subtracted before any interpolations have taken place. Briefly, this is done as follows. First a traditional reduction is done, so that a wavelength calibration is available. Second this wavelength calibration is used to determine the wavelength of each pixel in the uninterpolated frame. Third a function (cubic spline in $x$ and polynomial in $y$) is fitted to the original data values in the irregular grid of wavelength and $y$-coordinate. Fourth this model sky is subtracted from the frames, leaving only photon noise where the skylines have been subtracted. And fifth the frames are wavelength calibrated (interpolated) in the same way as in the traditional subtraction, but without any sky lines present.
Since we are only interested in the limited region around the emission line, we manually inspected and cleaned cosmic rays only around the line itself. A cosmic ray on the peak of the emission line of the first spectrum was cleaned by interpolating over the neighboring pixels.  A new wavelength solution was obtained by identifying the numerous sky lines throughout the spectrum. The identification of sky lines was done using the UVES sky atlas \citep{Hanuschik2003}. Finally, the new wavelength solution was applied to the stacked sky subtracted spectrum.

\begin{figure*}
  \centering
  \epsscale{1.0}
  \plotone{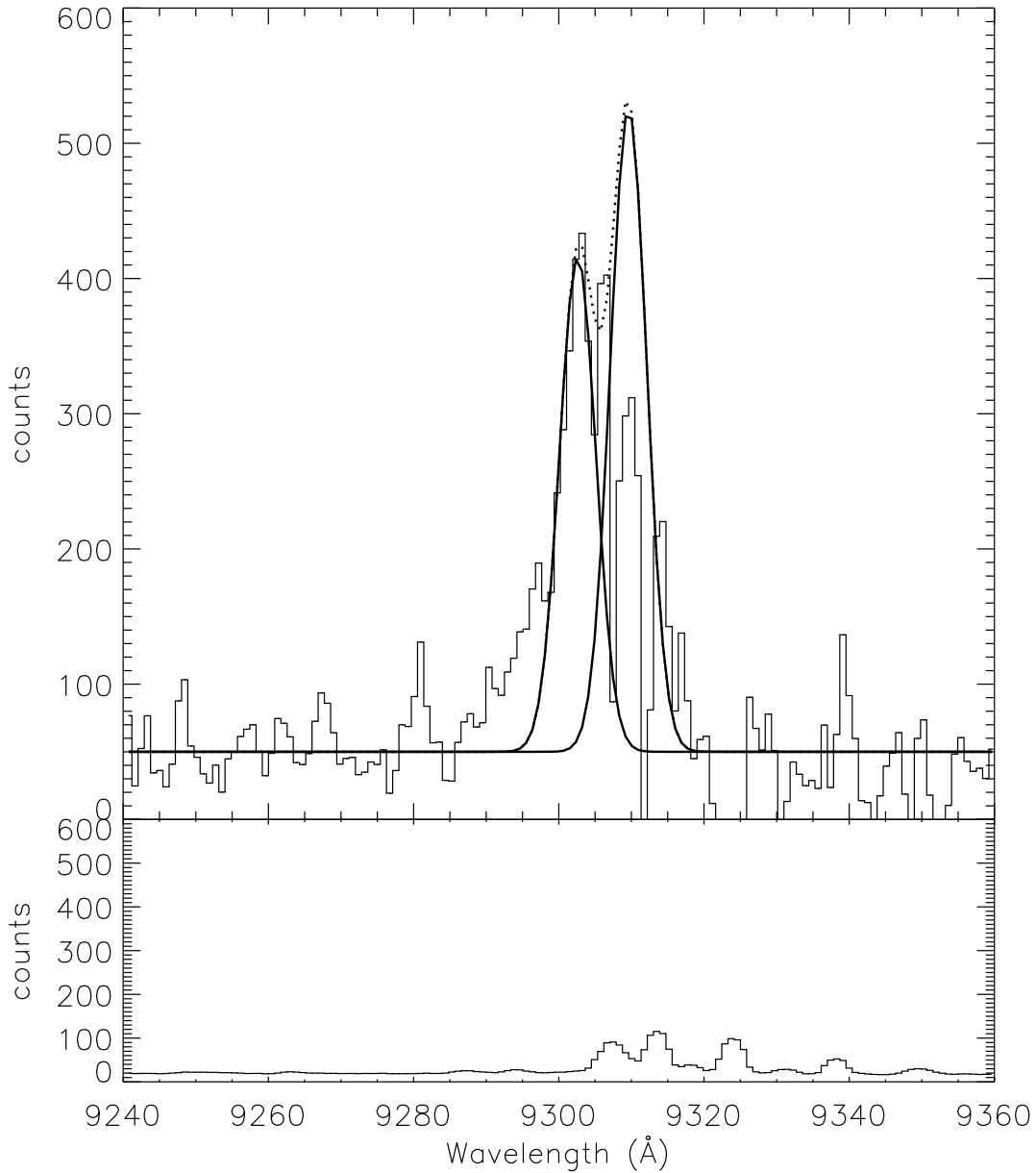}
  \figcaption[f4]{The [O\,{\scriptsize II}] doublet $\lambda3726.1, 3728.8$ {\AA} at $z=1.4959$ of GRB~070306 observed on 2007 May 7 using the VLT/FORS2 1028z-grism. The red component is affected by skyline residuals in the 9306 -- 9324{\AA}\ wavelength range and this is evident when comparing the spectrum to the prominent features in error spectrum. A Gaussian function with the same width as the LSF was fit to the blue component and the red component was assumed to be 30\% brighter than the red and the peaks separated by 7 \AA. The result is shown by the solid lines and the combined profile by the dotted line. \label{fig:1028z_spec}}
\end{figure*}

The line spread function (LSF) is estimated from the sky lines, giving FWHM $\sim 2.5$ {\AA}, while the detected emission covers approximately 20 {\AA} from 9295 to 9315 {\AA}. If the emission line is due to the [O\,{\scriptsize II}] doublet then a peak to peak separation of nearly 7 {\AA} is expected at this redshift. We fit a Gaussian function to the blue component of the doublet with a fixed width corresponding to the LSF. The red component, which is severely affected by the skyline residuals, is assumed to be 30\% brighter than the blue component and separated by 7 \AA. Fig.~\ref{fig:1028z_spec} shows the spectrum of the doublet with the profiles representing the unperturbed emission lines. The red component of the [O\,{\scriptsize II}] doublet (vacuum rest-frame wavelengths $\lambda3727.09, 3729.88$ {\AA}) is observed at  wavelength $9302.6 \pm 0.2${\AA}, which corresponds to a redshift of $z=1.49594 \pm 0.00006$.

The flux calibrated 300V spectrum was normalized to the $R$ band photometry. The integrated flux of the emission line was measured to be $(1.3 \pm 0.2) \times 10^{-16}$ erg s$^{-1}$ cm$^{-2}$. As this emission is significantly affected by the sky line residuals, we use our simple model to estimate a correction factor of 1.3 (by taking the ratio of the model and the spectrum). This is used to estimate the true flux of the [O\,{\scriptsize II}] doublet, which becomes $f_{\rm [O\,{II}], obs} = (1.7 \pm 0.6) \times 10^{-16}$ erg s$^{-1}$ cm$^{-2}$. Note that due to the uncertainty in the flux correction we have added a  systematic error. The star formation rate (SFR) can be computed using the relation to [O\,{\scriptsize II}] luminosity given by \citet[][Eq. 3]{Kennicutt1998},
$${\rm SFR}_{\rm [O\,{II}]} = (1.4 \pm 0.4) \times 10^{-41} L_{\rm [O\,{II}]} \ {\rm M}_\odot\, {\rm yr}^{-1} = 34^{+25}_{-18} \ {\rm M_\odot} {\rm yr}^{-1} \ ,$$
where $L_{\rm [O\,{II}]} = 4 \pi d_l^2 f_{\rm [O\,{II}], obs}$ and $d_l$ is the luminosity distance.

\placefigure{fig:1028z_spec}

\section{The host galaxy}
\label{sec:host}

With an almost featureless optical spectrum, the host galaxy at first seems difficult to model. In the optical we have no indication of an afterglow contribution (variability) and the magnitudes are therefore taken as measurements of the host. The late time NIR photometry is also interpreted as a measure of the NIR host properties. This gives a total of 10 independent magnitudes; SDSS/$ugriz$, FORS/$R$, NOT/$I$ and LIRIS/$JHK$. The adopted host magnitudes are listed in Table~\ref{tab:host}.

The optical colors are consistent with the featureless spectrum, in that they seem to represent a nearly flat continuum. This trend does not continue at the very red end of the spectrum and in the NIR bands. The SDSS $z$ band deviates from the other bluer bands, although with a large error, and the emission in the NIR peaks in the $H$ band. We investigate the SED properties by exploring the possible significance of typical emission lines such as H\,$\alpha$, H\,$\beta$, [O\,{\scriptsize II}] and [O\,{\scriptsize III}]. The emission-line strengths depend primarily on star burst activity (star formation rate), the age of the stellar population, extinction and metallicity. 
The host SED is modeled using the new\footnote{\url{http://www.ast.obs-mip.fr/users/roser/hyperz/}} Hyper-Z \citep{hyperz} (v11) along with four spectral synthesis models based on \citet[][hereafter CWW]{Coleman1980} and two empirical templates \citet{Kinney1996}. The Hyper-Z code convolves the templates with the filter transmission curves to obtain colors which are compared to those observed by means of a minimum chi-squared. This provides some constraint on the host extinction, qualitative information on the SED and an independent constraint on the redshift (which for this reason is left as a free parameter). The CWW templates lack detailed emission features,  so the \citet{Kinney1996} star burst templates (SB1, SB2) are crucial in order to check their relevance in explaining the observed SED. The difference between the two \citet{Kinney1996} templates is primarily the amount of internal extinction. For the SB1 template $E(B-V) < 0.1$ and for SB2 $0.11 < E(B-V) < 0.21$. A satisfactory fit was achieved for all extinction curves we explored, but the MW \citep[][hereafter S79]{Seaton1979} curve gave the best result with the \citet{Kinney1996} SB2 template and a relatively well constrained redshift $z = 1.55_{-0.26}^{+0.17}$ and a $\chi^2/{\rm dof} = 2.3/9$. The chi-squared values were all admittedly low, probably suggestive of the errors being color large (possibly due to the SDSS errors).

\begin{figure*}
  \centering
  \epsscale{1.0}
%  \plotone{f5.eps}
\includegraphics[angle=90, width=16cm]{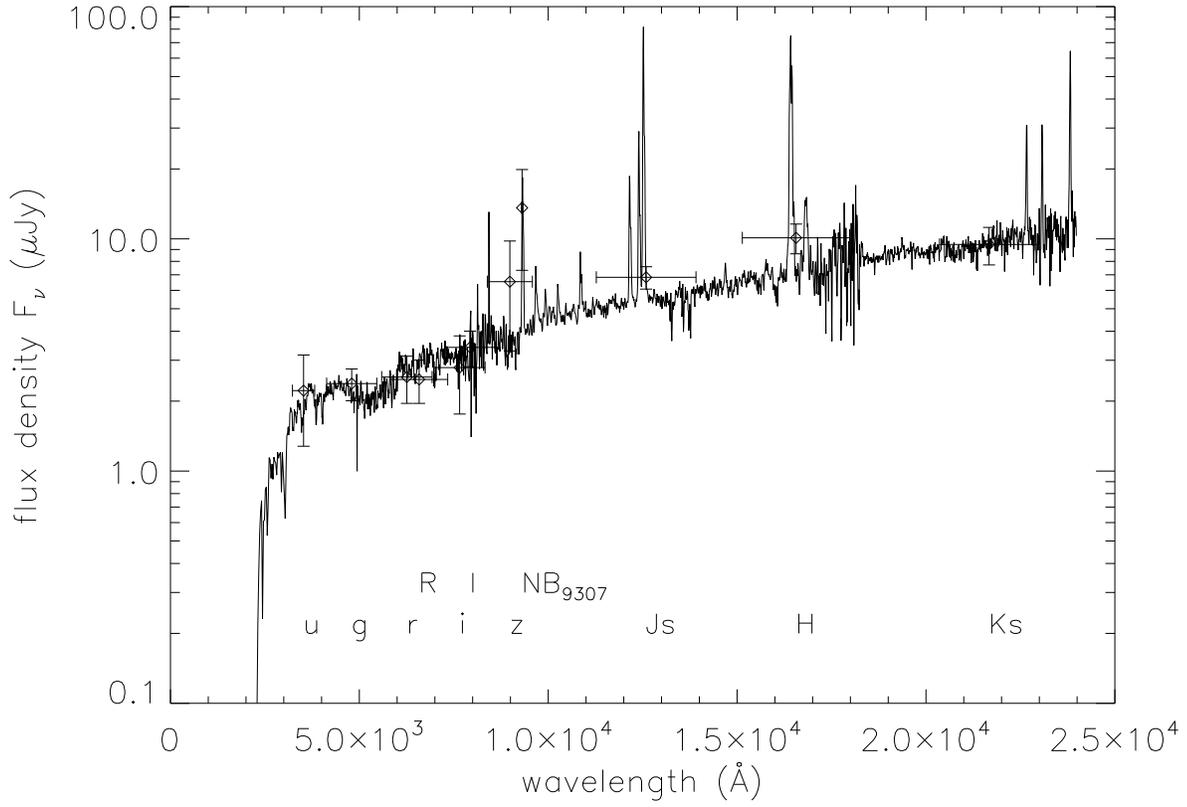}
  \figcaption[f5]{Best fitting SED ($\chi^2=2.5/10$) using Hyper-Z and a \citet{Kinney1996} type SB2 template at redshift $z=1.50$ with a MW-type extinction curve \citep{Seaton1979} and $A_V = 0.17$. The fit is based on 10 optical and NIR broad-band filters and an additional pseudo narrow-band filter NB$_{9307}$. \label{fig:NBsed}}
\end{figure*}

From the spectroscopic results we have a firm constraint on the emission line and this can be implemented in the SED fitting as a pseudo narrow-band measurement at the wavelength of the detected emission line. The NB$_{9307}$ filter has an effective wavelength of 9307.5 {\AA} and a 19 {\AA} bandpass. The flux of the NB$_{9307}$ filter was determined from the flux calibrated spectrum, giving NB$_{9307} = 21.1 \pm 0.5$ mag (AB). Running Hyper-Z on the 11 magnitudes resulted in a best fit redshift of $z=1.50 \pm 0.01$ and $\chi^2/{\rm dof} = 2.5/10$ and the same combination of the MW (S79) extinction curve and SB2 template. The resulting SED is shown in Fig.~\ref{fig:NBsed}. We thus conclude that the redshift of the host of GRB~070306 is $z = 1.4959$, as deduced from spectroscopy. Results from the host SED fits are summarized in Table~\ref{tab:hostextfits}.

%In Fig.~\ref{fig:NBsed} we can see that most of the filters are compatible with the SED continuum slope apart from the SDSS $z$-, $J$- and $H$ bands. This deviation is compatible with the suggestion that the filters cover the emission lines of [O\,{\scriptsize II}], H\,$\beta$+[O\,{\scriptsize III}] and H\,$\alpha$, respectively.
The broad band photometry fits to the SED templates do not indicate significant reddening. For the various extinction laws we find the upper limit of $A_V$ in the range $0.1 -- 0.5$. For the best fitting (SB2) template we find an upper limit of $A_V \le 0.45$.
 
At the host redshift the [O\,{\scriptsize II}] emission occurs in the range $\lambda$ 9304--9310 {\AA}, near the edge of the 1028z and 300V grisms. The emission line of [Ne\,{\scriptsize III}] is expected at 9660 {\AA} at this redshift, but unfortunately falls outside the range of both grisms. The best option to independently verify the host redshift is therefore to obtain low resolution spectroscopy in $J$ and $H$ bands to search for the [O\,{\scriptsize III}],  H\,$\beta$ and H\,$\alpha$ emission lines.

Our best fitting SED template is a star burst galaxy template. This is typical among GRB host galaxies \citep{Christensen2004}. The UV continuum flux in a star burst galaxy is dominated by the young (bursting) star population, and hence an appropriate measure of the star formation rate. We compute the SFR from the rest frame UV flux at 2800 {\AA}. The UV flux density is measured at  the observed $2800 (1+z)$ {\AA} wavelength in the flux calibrated spectrum and gives $f_{\rm 2800, obs} = 3.0 \pm 0.2\,  \mu$Jy, which corresponds to $L_\nu = 4\pi d_l^2 f_{\rm 2800, obs} / (1+z) = (17.1 \pm 1.2) \times 10^{28}$ erg s$^{-1}$ Hz$^{-1}$. Using \citet[][Eq.\ 1]{Kennicutt1998}, which relates the SFR to the luminosity at $\lambda2800$, we obtain SFR$_{\rm 2800} = 24 \pm 2 \ {\rm M}_{\odot}$ \, yr$^{-1}$. The error on the SFR does not include the roughly 30\% error of the Kennicutt relation. The normalized luminosity $L/L^*$ is estimated from the rest frame absolute B magnitude, $M_B = -22.3$, which is computed by Hyper-Z. Adopting $M^* = -21.0$ as the characteristic luminosity of field galaxies we find a normalized luminosity for the host of $L/L^* = 3.3$. This is bright in comparison to other GRB hosts, which primarily are sub-luminous ($L<L^*$) galaxies \citep[e.g.][]{Fruchter2006}. The luminosity weighted star formation rate sSFR $= 7.3$ M$_\odot$ yr$^{-1} (L/L^*)^{-1}$. A similar UV estimate at a rest frame wavelength of 1500 {\AA} can be computed using \citet{Madau1998} and the observed flux density $f_{\rm 1500, obs} = 1.8 \pm 0.6\ \mu{\rm Jy}$. This gives $L_\nu = (10.3 \pm 3.5) \times 10^{28}\ {\rm erg}\ {\rm s}^{-1}\ {\rm Hz}^{-1}$ and an SFR$_{1500} = 14.4 \pm 4.9\ {\rm M}_\odot$ yr$^{-1}$. 
%Note that the relative error is much larger in this latter measurement because it was  deduced from the SDSS photometry, whereas the measurement of $f_{2800}$ came from using the VLT $R$ band observation. 
These SFR estimates are consistent with each other (bearing in mind the 30\% uncertainty of the SFR relation) and with the SFR([O\,{\scriptsize II}]) computed in Sec.~\ref{sec:spec}. However, the SFR values are higher than most of the GRB host galaxies studied in e.g.\   \citet{Christensen2004}.

\section{Afterglow and extinction}
\label{sec:afterglow}

Variability in GRB~070306 was established early on thanks to the WHT/LIRIS data, where a brightening in the $K$ band of a factor two was detected during the first night \citep{gcnc6174,gcnc6176}. Further observations using VLT/ISAAC on the following two nights confirmed the $K$ band variability, and revealed variability in the $H$ band. The complete $JHK$ light curve is shown in Fig.~\ref{fig:lightcurve}. The variability of the afterglow component is most dramatic in the $K$ band and less pronounced in the $H$ band. In the $J$ band only marginal variability, if any at all, is visible during the first night. An afterglow was thus detected in the $H$ and $K$ bands during the first three nights.

As described in Section~\ref{sec:phot} we have used image subtraction to measure the afterglow flux, and if undetected set upper limits to its contribution. To establish the SED of the afterglow, we computed the semi-simultaneous flux of the afterglow based on the observations carried out on the second night between 2007 Mar 08.054 UT and 08.070 UT. The ISAAC data in this period are superior in quality and therefore best suited for a contemporaneous SED analysis in addition to the FORS $R$ band observation at Mar 08.104 to characterize the SED.
We used the AB-offsets calculated by Hyper-Z for the $R$,$I$,$J$,$H$ and $K$ filters and found the values 0.23, 0.45, 0.94, 1.41 and 1.87, respectively.\label{ABoffs} 

\begin{figure*}
  \centering
  \epsscale{1.0}
  \plottwo{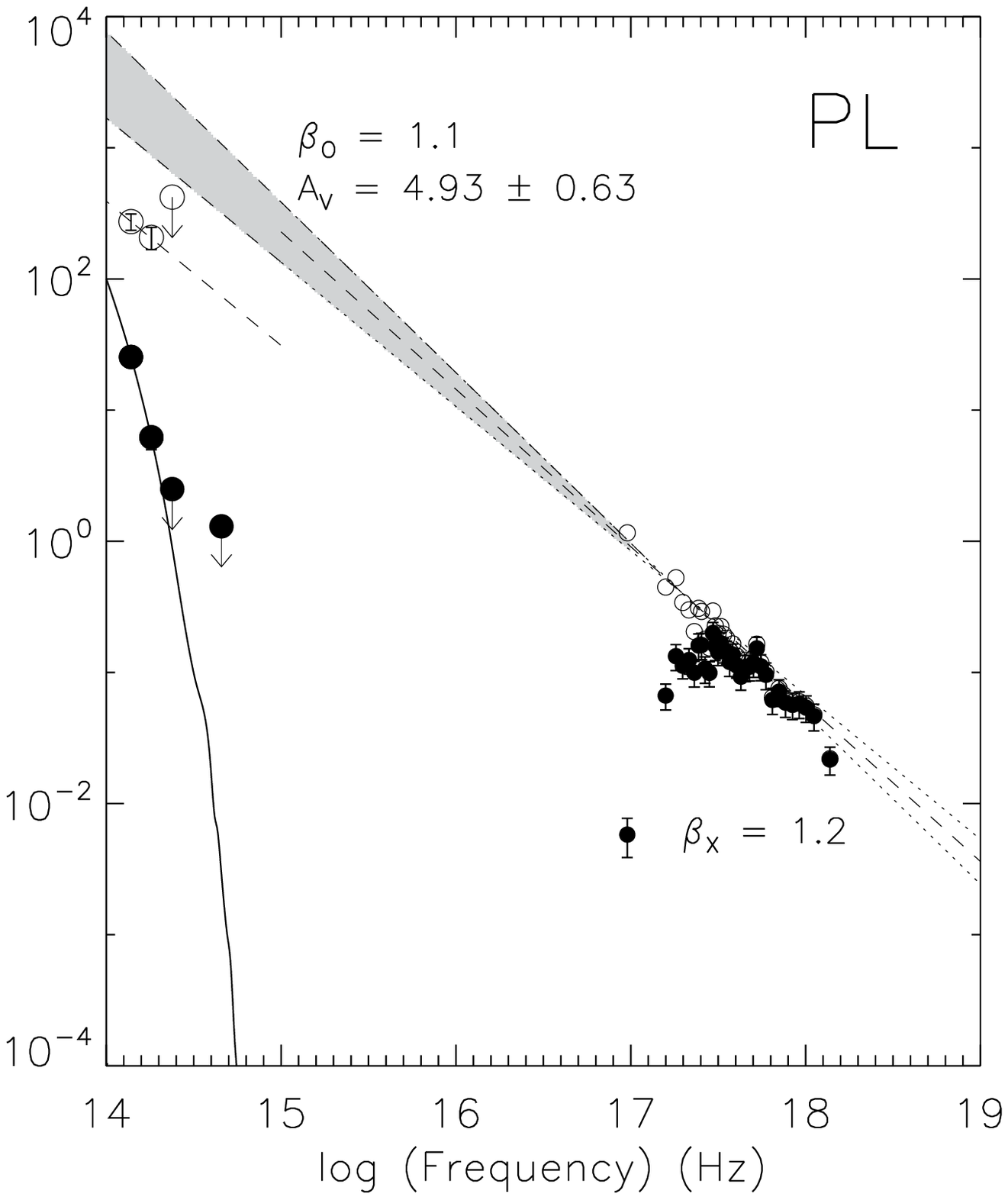}{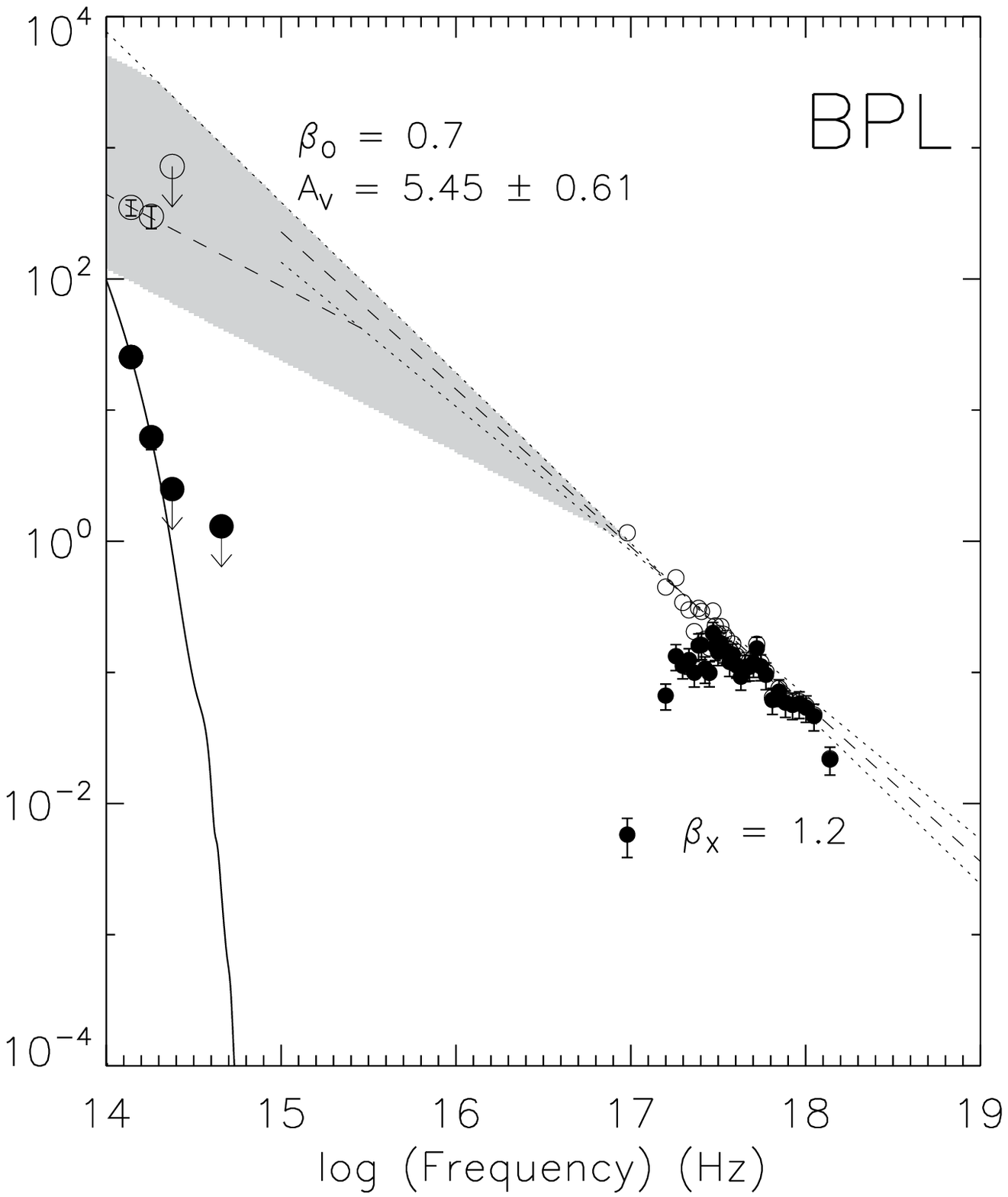}
  \figcaption[f6]{Various extinction curves are fit to the optical and NIR afterglow measurements at $t_0 + 1.38$ days by assuming synchrotron emission in the form of a single power-law (PL) or broken power-law (BPL) spectral slope. A separate power-law fit to the XRT data gave a spectral slope of $\beta_X = 1.2 \pm 0.1$, which is used to constrain the optical/NIR slope, $\beta_O$. The best fit is obtained using the SMC extinction-curve of \citet{Pei1992} (represented by the restframe visual extinction, $A_V$). The observed optical/NIR measurements are shown as large filled circles, whereas the large open circles represent the same data corrected for the best fit extinction. The small filled dots with error bars represent the XRT data corrected for a fixed Galactic absorption, while the small open dots represent the same data corrected for a variable equivalent hydrogen column density. The shaded areas represent the limits on the total extinction without assuming an extinction curve (see text). The single power-law model (left) is not consistent with the data for this \citet{Pei1992} SMC curve, but is marginally consistent using the \citet{Prevot1984,Bouchet1985} SMC curve.
  \label{fig:agfit}}
\end{figure*}

With an afterglow detection in only two bands and two additional upper limits  there is a clear degeneracy in estimating the extinction and spectral slope. We need to introduce constraints on the spectral index and one option is to assume that the optical spectral index is identical to the X-ray, i.e. $\beta_O = \beta_X$. We fit a number of extinction curves (see Table~\ref{tab:hostextfits}) to the broad band optical/NIR data and compare the fits in terms of chi-squared (see Table~\ref{tab:extfits}). The fit is made by assuming an intrinsic (unobscured) spectral slope and fitting the amount of reddening ($A_V$) required such that the corrected points match the assumed spectral slope. The resulting $A_V$ values are very similar and mostly consistent with each other within the errors. It is difficult to discriminate between the various extinction curves and the MW and LMC curves give the same results, but the SMC extinction curve of \citet{Pei1992} (R$_V = 2.93$) does provide a slightly better fit than the others. By requiring that the NIR slope matches the hardest X-ray slope ($\beta_O = 1.1$), we obtain an $A_V = 4.9 \pm 0.6$. However, the extinction corrected $H$ and $K$ points do not match the extrapolation of the X-ray power-law and a single power-law model is therefore excluded by this fit. If the errors are increased by a factor of three then a marginally consistent result can be achieved. The only extinction curve which does allow for the PL model to succeed is the \citet{Pei1992} SMC type curve.
In the context of the synchrotron model, there is an additional possibility for the optical/X-ray spectrum, namely a broken power law with the cooling break between the two bands.
This requires that $\beta_O = \beta_X - 0.5$ and gives a spectral slope of $\beta_O = 0.7 \pm 0.1$ using the $\beta_X$ value from Sec.~\ref{sec:xrt}. The best fitting extinction curve is obtained using $A_V = 5.4 \pm 0.6$. The errors represent the uncertainty related to the observational errors, which are dominated by the uncertainty of the $H-K$ color. The SED and extinction curve fits are shown in Fig.~\ref{fig:agfit}.
%The cooling frequency is not significantly constrained by the data, other than by the implication that the cooling break is between the optical and X-ray in the case of a broken power law.

The constraints on the total extinction, $A_\lambda$, by exploring the possible X-ray spectral slopes and cooling break locations under the assumption that one of the two power-law models is correct. In the case of a single power-law we require that the extinction is not less than the shallowest X-ray slope supported by the data, ie.\,$\beta > 1.1$ .
Similarly, for the broken power-law model we use the maximum possible cooling break frequency to determine the minimum total extinction. The maximum total extinction comes from the limit of the minimum cooling break frequency, which is set by the frequency of the bluest band with an optical/NIR afterglow detection (eg.\ the $H$ band). The results are given in Table~\ref{tab:totext} and are visually indicated by the shaded regions in Fig.~\ref{fig:agfit}.

\section{Summary and discussion}
\label{sec:summary}

The observational data of GRB~070306 presented here suggests that this is a dark burst with a very red NIR afterglow and that the host ($z=1.4959$) is brighter than typical host galaxies \citep{Berger2007}. GRB~070306 possesses a higher than normal star formation activity in comparison to other GRB hosts. The afterglow SED analysis shows that significant reddening is required to explain the observed optical/NIR afterglow assuming typical power-laws between the optical/NIR and X-rays. From the combined constraints of the X-ray and optical/NIR data and the assumption of a single or broken power-law, we can set limits on the total extinction without requiring a specific extinction law. We find that a single power-law is not supported by the data for most of the extinction laws explored, while a broken power-law and $A_V = 5.4 \pm 0.6$ \citep[SMC, ][]{Pei1992} best describes the observed colors.

The Galactic gas-to-dust ratio \citep{Predehl1995} corresponds to $N_H = 1.7\times 10^{21} {\rm cm}^{-2}/A_V$ (calibrated using X-ray absorption). 
Using the same ratio, the observed X-ray $N_H$ would correspond to $A_V = 15$, which is much larger than observed, indicating a lower dust-to-gas ratio. Similar results have been found in the past, attributing the apparent deficit to dust destruction, dust depletion, or an unusual ISM composition \citep{Galama2001,Watson2006,Campana2007}.
%While the latter recent discoveries suggest metal-lines may not originate near the progenitor \citep[$\sim$kpc][]{Vreeswijk2007,Prochaska2007}, this is still unsettled.

In the case of GRB~070306, the combination of what seems to be a highly obscured (and reddened) afterglow and a host with little or no indication of reddening illustrates the general concept that GRB progenitors reside in particularly dense regions within their host galaxies. If, by chance, the GRB line-of-sight traverses large columns of dust and gas while the host itself appears unobscured, this suggests the dust is unevenly distributed and likely concentrated in smaller regions or clouds within the host galaxy. Such patchy patterns are in effect also seen in our own Galaxy, although it does not necessarily imply that dust and gas trace each other. \citet{Padoan2006} find evidence supporting the latter and suggests spatial fluctuations of the dust-to-gas ratio, even on very small scales, exist in the MW. 
If the dust obscuration occurs in smaller regions with supposedly varying optical depths, this may also have implications on absorption line studies and the metallicities resulting from them. There are several element abundance studies in the literature indicating dust depletion \citep[e.g.\ ][]{Savaglio2004,Watson2006}, but few (if any) cases providing a robust comparison of emission and absorption line induced metallicities. Such studies may therefore offer some hope in providing a better understanding of the isolated and global properties of these galaxies.

Highly extinguished GRB afterglows are rare, although a few discoveries implying dust have been reported in the past \citep{Masetti2000,Klose2000} and more recently \citep{Levan2006,Rol2007,Castro-Tirado2007,Tanvir2007}.  Early NIR observations are crucial in detecting the afterglow of dark bursts, and GRB~070306 illustrates the importance of following NIR afterglows with $8--10$m class telescopes. With the increasing number of apparent dust obscured (very red) afterglows, all of which are classified as dark, we may be nearing an understanding of the fairly large fraction of dark GRBs. The nature of gamma-ray bursts and their association with regions of intense star-formation, and accordingly an abundance of gas and dust, provides a natural dust interpretation for dark bursts within the context of the synchrotron model. Although there are alternative explanations, the dust-related interpretation seems intuitively inline with the observed red color, the range of redshifts and host properties. An alternative is that the X-ray and optical afterglows have separate physical origins \citep{Perley2008}. As the number of well observed dark bursts increase we are likely to discover many more of these systems, and hopefully settle the ambiguities of interpretation.


%% Included in this acknowledgments section are examples of the
%% AASTeX hypertext markup commands. Use \url without the optional [HREF]
%% argument when you want to print the url directly in the text. Otherwise,
%% use either \url or \anchor, with the HREF as the first argument and the
%% text to be printed in the second.

\acknowledgments

We would like to thank the referee for a thorough review and helpful suggestions, which no doubt has improved the manuscript. 
AOJ acknowledges support from the Norwegian Research Council. ER acknowledges support from STFC.
The Dark Cosmology Centre is funded by the DNRF. The research activities of JG are supported by the Spanish Ministry of Science through the programmes ESP2005-07714-C03-03 and AYA2004-01515.

\bibliographystyle{apj}
\bibliography{mnemonic,paper_cites}

\clearpage

\begin{deluxetable}{lrlllrrrrr}
\tabletypesize{\small}
\rotate
\tablewidth{20cm}
\tablecaption{Observing log and results from photometry \label{tab:obslog}}

\tablehead{
\colhead{Date} & \colhead{Time\tablenotemark{a}} & \colhead{Instrument}  & \colhead{Band}  & \colhead{Seeing} & \colhead{T$_{\rm int}$} & \colhead{Magnitude\tablenotemark{b}}  & \colhead{Error} & \colhead{F$_\nu$\tablenotemark{c}} & \colhead{Error} \\
\colhead{UT} & \colhead{days} & & &  & \colhead{s} & \colhead{Vega} & \colhead{mag}  & \colhead{$\mu$Jy} & \colhead{$\mu$Jy} \\
}

\startdata
Mar 06.8889  & 0.191 &  WHT / LIRIS &  J & 1\farcs1 & 1200 & $>22.4$ & \dots & $<1.7$ & \dots\\
Mar 08.0542 & 1.357 & VLT / ISAAC &  J  & 0\farcs9 & 1040 & $>22.0$ & \dots & $<2.5$ & \dots \\
Mar 09.2342 & 2.537 & VLT / ISAAC & J & 0\farcs7 & 2560 & (20.90)$^\dagger$ &  0.13 & 6.82 & 0.75 \\
Apr 30.8871  & 55.190 & WHT / LIRIS & J & 0\farcs8 & 1800 & $--$ & \dots & \dots & \dots \\
Mar 06.8670  & 0.170 & WHT / LIRIS &  H & 1\farcs1 & 1200 & 19.78 & 0.12 & 12.31 & 1.36 \\
Mar 08.0836  & 1.386 & VLT / ISAAC & H & 0\farcs8 & 1152 & 20.52 &  0.21 & 6.23 & 1.20 \\
Mar 09.2074  & 2.510 & VLT / ISAAC & H & 0\farcs6 & 1344 & (20.00)$^\dagger$ &  0.20 & 10.05 & 1.48 \\
Apr 30.9242  & 55.227 & WHT / LIRIS & H & 1\farcs0 & 1800 & $--$ & \dots & \dots & \dots \\
Mar 06.8416  & 0.144 &  WHT / LIRIS & K & 1\farcs0 & 1500 & 18.28 &  0.15 & 31.92 & 4.41 \\
Mar 07.0361  & 0.339 & WHT / LIRIS & K & 0\farcs9 & 1200  & 17.61 &  0.14 & 59.16 & 7.62 \\
Mar 08.0701  & 1.373 & VLT / ISAAC & K & 0\farcs7 & 1152 & 18.53 &  0.15 & 25.35 & 3.50 \\
Mar 08.2153 & 1.512 & VLT / ISAAC & K & 0\farcs6 & 1152 & 18.50 &  0.15 & 26.06 & 3.60 \\
Mar 09.1879  & 2.490 & VLT / ISAAC & K & 0\farcs6 & 1152 & 19.83 &  0.19 & 7.66 & 1.34 \\
Apr 30.9699  & 55.272 & WHT / LIRIS & K & 0\farcs9 & 1800 & $--$ &  \dots & \dots & \dots \\
Jan 30.2244  & 329.526 & VLT / ISAAC & K & 0\farcs5 & 1920 & (19.60)$^\dagger$ &  0.19 & \dots & \dots \\
\hline \\
Mar 06.9523  & 0.255 & NOT / StanCam & I & 1\farcs0 & 1200 & $--$ & \dots & \dots & \dots \\
Mar 12.0615  & 5.364 & NOT / ALFOSC & I & 0\farcs6 & 3600 &  (22.15)$^\dagger$  & 0.19 & 3.57 & 0.62 \\
Mar 08.1042  & 1.407 & VLT / FORS2 &  R  & 0\farcs9 & 120 & $>23.5$  & \dots & $>1.3$ & \dots\\
Apr 20.0507  & 44.364 & VLT / FORS2 &  R  & 0\farcs9 & 180 & $>23.5$  & \dots & $<1.3$ & \dots \\
May 07.0040  & 61.306 & VLT / FORS2 &  R & 0\farcs7 & 60 & (22.78)$^\dagger$ & 0.23 & 2.53 & 0.54 \\
\enddata
\tablenotetext{a}{Time since burst trigger t$_0$ = 2007 Mar 06.6976 UT}
\tablenotetext{b}{The photometry was obtained by measuring the flux in circular apertures of 3\farcs0 radius (using SExtractor) on the best data available (around Mar 09.2, t=1.38 days and best in terms of sub-arcsecond seeing, photometric conditions and telescope aperture). All other magnitudes and upper limits are measured from image subtracted frames and calibrated using the reference frames, which are identified by the daggers (see Sec.~\ref{sec:phot} for details).}
\tablenotetext{c}{The flux values have been computed using the AB offsets (see Section~\ref{ABoffs}) and correcting for Galactic extinction according to \citet{Schlegel1998}}

\end{deluxetable}

\begin{deluxetable}{lrrrrrrrrrrr}
\tabletypesize{\small}

\tablecaption{Host magnitudes}

\tablehead{
\colhead{} & \colhead{u} & \colhead{g} & \colhead{r} & \colhead{i}  & \colhead{z}  & \colhead{R}  & \colhead{I} & \colhead{NB$_{9307}$} & \colhead{J} & \colhead{H} & \colhead{K} 
}

\startdata
Mag(AB) & 23.17  & 23.06 &  22.96 & 22.84 & 21.90 & 22.98 & 22.62 & 21.1 & 21.84 & 21.41 & 21.47 \\
Error & 0.46 & 0.17 & 0.25 & 0.40 & 0.54 & 0.23 & 0.19 & 0.5 & 0.12 & 0.16 & 0.20 \\
\enddata

\tablecomments{The SDSS $ugriz$ colors are provided by \citet{gcnc6170}. See Section~\ref{sec:host} for a description of the pseudo narrow-band NB$_{9307}$.\label{tab:host}}

\end{deluxetable}

\begin{deluxetable}{lrrrrrrr}
\tabletypesize{\small}

\tablecaption{Results from the GRB~070306 host galaxy SED fit of ten broad-band filters and one pseudo narrow-band filter using Hyper-Z. The extinction curves used here correspond to those implemented in Hyper-Z, in addition to \citet{Pei1992} for the Small Magellanic Cloud (SMC); \citet{Allen1976} for the Milky Way (MW), \citet{Seaton1979} fit by \citet{Fitzpatrick1986} for the MW, \citet{Fitzpatrick1986} for the Large Magellanic Cloud (LMC), \citet{Prevot1984} and \citet{Bouchet1985} for the SMC and \citet{Calzetti2000} for starburst galaxies (SB). \label{tab:hostextfits}}

\tablehead{
\colhead{Ext. law} & \colhead{SED} & \colhead{z} & \colhead{$\chi^2_\nu$} & \colhead{$A_V$}  & \colhead{max($A_V$)\tablenotemark{a}} & \colhead{$z_{\rm low}$} & \colhead{$z_{\rm up}$} \\
}

\startdata
MW (A76)            & SB2 & 1.49 & 0.256 & 0.17 & 0.43 & 1.49 & 1.51 \\
MW (S79, F86)   & SB2 & 1.50 & 0.249 & 0.17 & 0.45 & 1.49 & 1.51 \\
LMC (F86)           & SB2 & 1.49 & 0.268 & 0.17 & 0.45 & 1.48 & 1.51 \\
SMC (P84, B85) & SB1 & 1.49 & 0.316 & 0.03 & 0.23 & 1.48 & 1.51 \\
SB (C00)             & SB2 & 1.49 & 0.269 & 0.23 & 0.47 & 1.48 & 1.51 \\
SMC (P92)          & SB1 & 1.49 & 0.325 & 0.00 & 0.08 & 1.48 & 1.51 \\
\enddata
\tablenotetext{a}{The ($1\sigma$) upper limit of $A_V$.}

\end{deluxetable}

\begin{deluxetable}{lrrrr}
\tabletypesize{\small}

\tablecaption{Rest-frame $A_V$ and $\chi^2$ (${\rm dof} = 1$) results from fitting the afterglow SED for various extinction curves. While the MW and LMC curves provided similar results, the overall best fitting curve was the SMC curve of \citet{Pei1992}. The data are not consistent with the (single) PL model, except in the case of the SMC curve of \citet{Prevot1984,Bouchet1985}, where a marginal consistency could be achieved. \label{tab:extfits}}

\tablehead{
 & \multicolumn{2}{c}{Power-law} & \multicolumn{2}{c}{Broken\- power-law} \\
 \cline{2-3} \cline{4-5} \\
\colhead{Ext. law} & \colhead{$A_V$}  & \colhead{$\chi^2$} & \colhead{$A_V$}  & \colhead{$\chi^2$} \\
}

\startdata
MW (A76) & 4.81 & 0.17 & 5.31 & 0.27 \\
MW (S79,F86) & 4.81 & 0.17 & 5.31 & 0.27 \\
LMC (F86) & 4.81 & 0.17 & 5.31 & 0.27 \\
SMC (P84, B85) &  6.00 & 0.65 & 6.71 & 1.13 \\
SB (C00) & 5.15 & 0.48 & 5.75 & 0.81 \\
SMC (P92) & 4.93 & 0.17 & 5.45 & 0.26 \\
\enddata

\end{deluxetable}

\begin{deluxetable}{lrrrr}
\tabletypesize{\small}

\tablecaption{Total extinction constraints. \label{tab:totext}}

\tablehead{
 & \multicolumn{2}{c}{Power-law} & \multicolumn{2}{c}{Broken power-law} \\
 \cline{2-3} \cline{4-5} \\
\colhead{Filter} & \colhead{min($A_\lambda$)}  & \colhead{max($A_\lambda$)} & \colhead{min($A_\lambda$)}  & \colhead{max($A_\lambda$)} \\
}

\startdata
$R$ & 6.0 & \dots & 3.8 & \dots \\
$J$ & 6.0 & \dots & 4.4 & \dots \\
$H$ &  5.4 & 6.9 & 2.8 & 6.9 \\
$K$ & 4.2 & 5.7 & 1.5 & 5.6 \\
\enddata

\end{deluxetable}

\clearpage

%% the file, uncomment the \plotone, \plottwo, and
%% If you need a layout that cannot be achieved with \plotone or
%% \plottwo, you can invoke the graphicx package directly with the
%% \includegraphics command or use \plotfiddle. For more information,
%% please see the tutorial on "Using Electronic Art with AASTeX" in the
%% documentation section at the AASTeX Web site,
%% http://www.journals.uchicago.edu/AAS/AASTeX.
%%
%% The examples below also include sample markup for submission of
%% supplemental electronic materials. As always, be sure to check
%% the instructions to authors for the journal you are submitting to
%% for specific submissions guidelines as they vary from

%% This example uses \plotone to include an EPS file scaled to
%% 80% of its natural size with \epsscale. Its caption
%% has been written to indicate that additional figure parts will be
%% available in the electronic journal.

%\begin{figure}
%\plotone{lightcurve.pdf}
%\caption{Multi-band lightcurve.\label{fig1}}
%\end{figure}

\clearpage

%% Here we use \plottwo to present two versions of the same figure,
%% one in black and white for print the other in RGB color
%% for online presentation. Note that the caption indicates
%% that a color version of the figure will be available online.
%%

%\begin{figure}
%\plottwo{f2.eps}{f2_color.eps}
%\caption{A panel taken from Figure 2 of \citet{rudnick03}. 
%See the electronic edition of the Journal for a color version 
%of this figure.\label{fig2}}
%\end{figure}

%% This figure uses \includegraphics to scale and rotate the still frame

%\begin{figure*}
%  \centering
%  \epsscale{1.0}
%  \plotone{f2.eps}
%  \figcaption[f2]{The smoothed FORS2 300V 2D-spectrum indicating the location of [O\,{\scriptsize II}] $\lambda3727$ at 9304 {\AA}.\label{fig:2d_spec}}
%\end{figure*}

\clearpage

\end{document}